\documentstyle[namedreferences,psfig]{spackap}

\def\kms{km~s$^{-1}$ }
\def\Lya{Ly$\alpha$ }

\def\Lyb{Ly$\beta$ }

\def\Lyg{Ly$\gamma$ }

\def\cm2{cm$^{-2}$ }
\def\etal{{\it et al.} }

\begin{opening}

\title{On the Measurements of D/H in QSO Absorption Systems}
\subtitle{Closing in on the primordial abundance of deuterium}

\author{Scott \surname{BURLES}}
\institute{Department of Astronomy \& Astrophysics, University of Chicago, 5640 S. Ellis Ave, Chicago, IL 60637}
\author{David \surname{Tytler}}
\institute{Center for Astronomy and Astrophysics, University of California, San Diego, 9500 Gilman Drive, La Jolla, CA 92093-0424}

\date{}

\end{opening}

\begin{document}

\begin{abstract}
We present our measurements of the deuterium to hydrogen
ratio (D/H) in QSO absorption systems, which give
D/H = 3.40 $\pm \, 0.25 \times 10^{-5}$ based on analysis of four independent
systems.  We discuss the properties of two systems which provide
the strongest constraints on D/H.
We outline the systematic effects involved in measurements of D/H 
and introduce a sophisticated method of analysis
which properly accounts for these effects.
\end{abstract}


\section{Introduction}
The status of standard big 
bang nucleosynthesis (BBN) has been extensively reviewed in these 
proceedings and in the recent literature
(Copi \etal 1995; Sarkar 1996; Hata \etal 1997; Cardall \& Fuller 1996;
Fuller \& Cardall 1996; Schramm \& Turner 1997).
It has been emphasized that a determination
of the primordial deuterium abundance has the potential to constrain
the predictions of BBN, test the standard model, and give a precise
measure of the present-day baryon density.  Recent measurements of D/H
towards bright, distant QSOs are now realizing this potential, and
are already giving improved constraints on the models of BBN.
(Tytler \& Burles 1997).

Hogan (1998) gives a thorough overview of the present status
of extragalactic D/H measurements; here we focus on measurements of
the systems for which we have obtained high-quality spectra with the
HIRES spectrograph (Vogt \etal 1994) and Keck1 10-m telescope.

\section{The Measurements}

We have analyzed four high-redshift QSO absorption systems
which place useful constraints on D/H.   We show the limits on D/H
in each of these systems in Fig. 1.  Two of the systems, towards
Q1251+3644 and Q1759+7539 have larger uncertainties for different reasons.
Q1251+3644 is a faint QSO (V=19), and over 10 hours of observing time
yields a spectrum with modest signal-to-noise ratio (SNR = 10) at 
high-resolution.  The SNR drops quickly at lower wavelengths, and we
cannot extract useful information from the highest order Lyman lines as
a result.  The system towards Q1759+7539 is a complex system of absorbers
with a very high neutral hydrogen column density, log N(H~I) $> 10^{19}$ 
(all column densities expressed in \cm2).
The analysis of that system gives only an upper limit on D/H, at the 
95\% confidence level; the
constraints are obtained from comparing the line profiles of \Lyb and \Lyg to
the profile of \Lya.  The method used to extract the limits on D/H 
will be described below.  These results are preliminary, and improved 
constraints will require more data to allow a robust analysis. 

\begin{figure}
\psfig{file=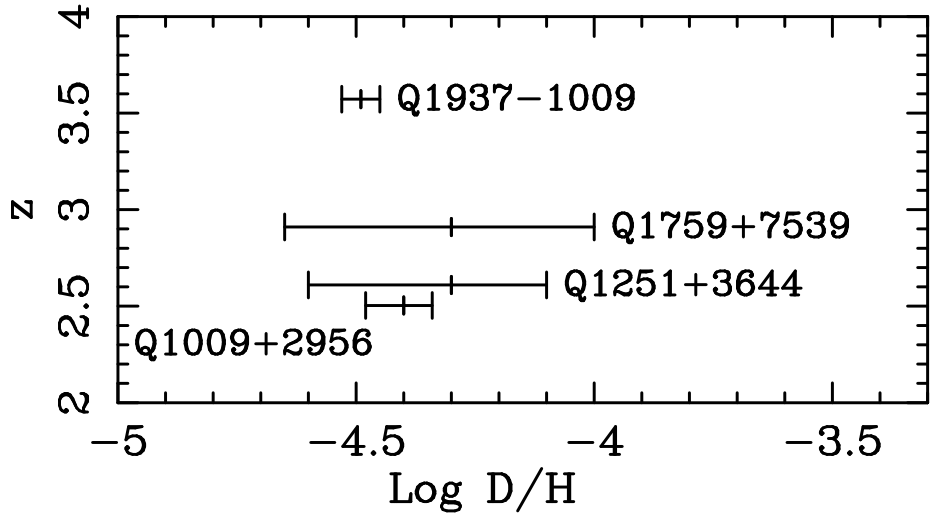,width=\textwidth}
\caption{Constraints on D/H from four independent measurements in 
four separate QSO absorption systems.  Shown are the most likely values
and 68\% confidence limits at the absorption redshift of the D/H systems.}
\end{figure} 

\begin{figure}
\psfig{file=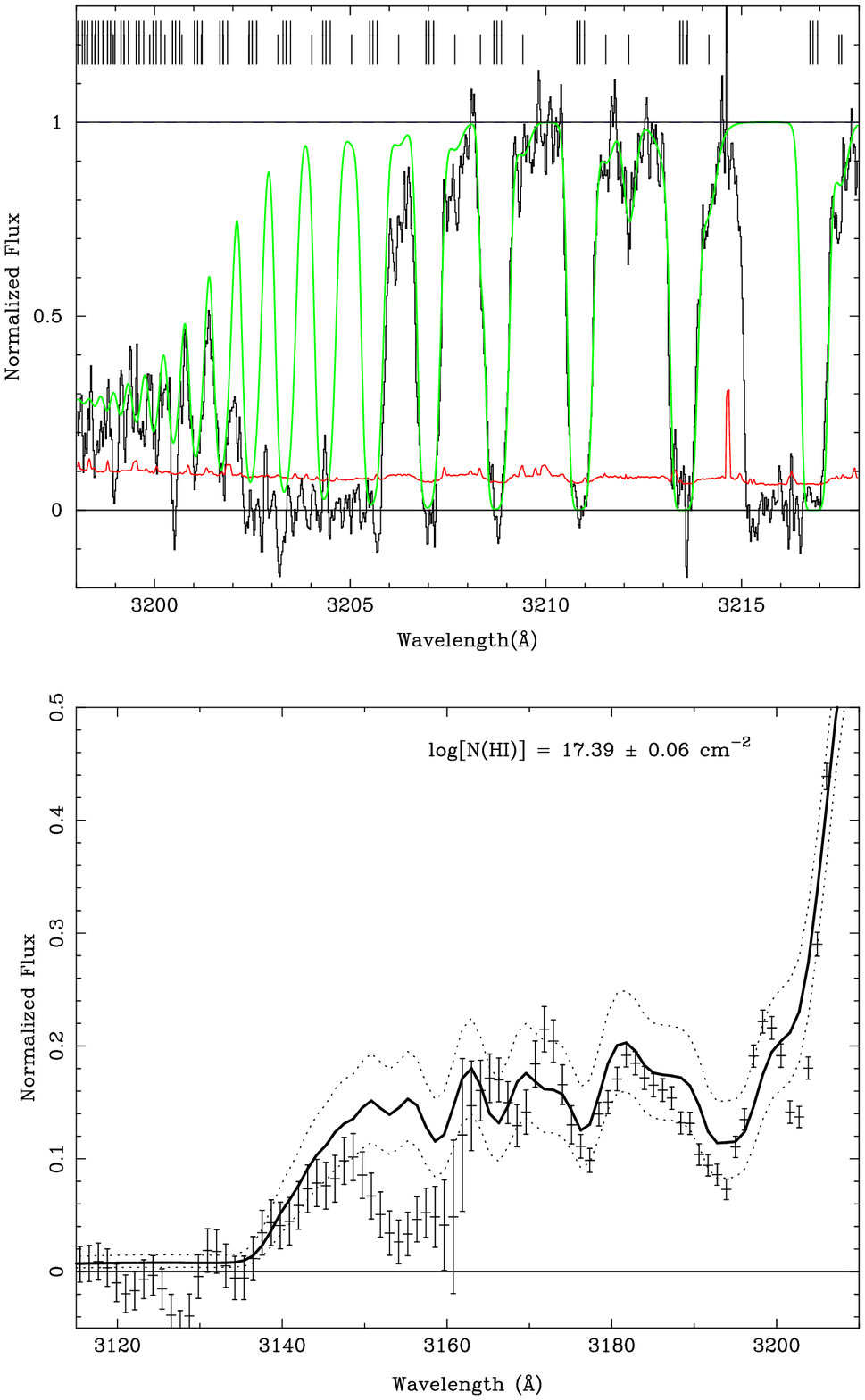,width=\textwidth}
\caption{Lyman limit region of Q1009+2956 at $z=2.504$.
\protect\\
{\it Top}: HIRES spectrum (FWHM = 8 km/s) showing Lyman lines 11 through 24,
and model fit with three components and total H~I column density, 
log N(H~I) = 17.31.
\protect\\
{\it Bottom}: Lick spectrum (FWHM = 4 \AA) showing residual flux and model
fit shortward of Lyman limit.  The solid line shows the best fit model with
log N(H~I) = 17.39, and dotted lines represent 68\% confidence levels.
Absorption shortward of 3135 \AA~is due to a lower redshift Lyman limit system.}
\end{figure}

Tytler \etal (1996) made the first measurement of low D/H in the absorption
system at $z=3.572$ towards Q1937--1009.  We analyzed the 
high-resolution spectrum (8 hrs of exposure), which resolved the entire Lyman
series up to Ly-19, as well as associated metal lines.  By profile
fitting the Lyman lines, with the position of the velocity components given
by the metal lines, we find D/H = 2.3 $\pm \, 0.3 \pm 0.3 \times 10^{-5}$
(statistical and systematic errors).  
The largest uncertainty in the measurement is the neutral hydrogen column
density, log N(H~I) = 17.94 $\pm \, 0.06 \pm 0.05$, 
and the uncertainty stems from the saturated Lyman profiles
(discussed in detail below).  We then obtained a high quality 
low-resolution spectra from Keck with LRIS (Oke \etal 1995), which gave better
sensitivity shortward of the Lyman limit, to directly measure the
total N(H~I) in the system and therefore place better constraints on D/H.
Utilizing both the high and low-resolution spectra, we find
log N(H~I) = 17.86 $\pm \, 0.02$ by a direct measurement of the optical depth
shortward of the Lyman limit at 4200 \AA~(Burles \& Tytler 1997a).
With this constraint and a more sophisticated fitting procedure, 
we measure D/H = 3.3 $\pm \, 0.3 \times 10^{-5}$ (Burles \& Tytler 1997b).

We discovered an absorption system at $z=2.504$ towards 
Q1009+2956 ideal for a measurement of D/H.  This system has a lower
hydrogen column density, log N(H~I) = $17.39 \pm 0.06$.  The
highest order Lyman lines become unsaturated, which yields a precise 
measurement of N(H~I) in both low and high resolution spectra (Fig. 2). 
Over twelve hours of Keck+HIRES produced a very high quality spectrum
of the entire Lyman series, resolving the entire series up to Ly-22.
We find strong evidence for contamination of the deuterium \Lya absorption
feature, which unfortunately introduces the largest uncertainty in the
measurement.  With the contamination included in the analysis, we measure
D/H = 4.0 $\pm \, 0.7 \times 10^{-5}$ (Burles \& Tytler 1997c).  

The four independent systems support a low primordial abundance of deuterium,
and together give D/H = 3.4 $\pm \, 0.25 \times 10^{-5}$.  
If this represents the primordial value, nucleosynthesis calculations
from standard BBN models with three light neutrinos give
the baryon-to-photon ratio, $\eta =  5.1 \pm \, 0.3 \times 10^{-10}$
and the baryon density, $\Omega_b\,h_{100}^2 = 0.019 \pm \, 0.001$.
We arrive at very small statistical errors (10\% at 95\% confidence),
so we must confront the existence of systematic effects.
In the remainder of this paper we present a sophisticated
method that we use in order to properly account for the systematic effects
currently known.

\section{An Improved Method to Measure D/H}

Many groups have reported detection of deuterium in absorption 
along the line of sight to distant
QSOs (Songaila \etal 1994; Carswell \etal 1994; Tytler \etal 1996,
Wampler \etal 1996; Rugers \& Hogan 1996a,b; Webb \etal 1997;
Burles \& Tytler 1997b,c).  
Here, we put forward an improved method to measure D/H,
which has many advantages over the current, widely used, methods.

\begin{figure}
\psfig{file=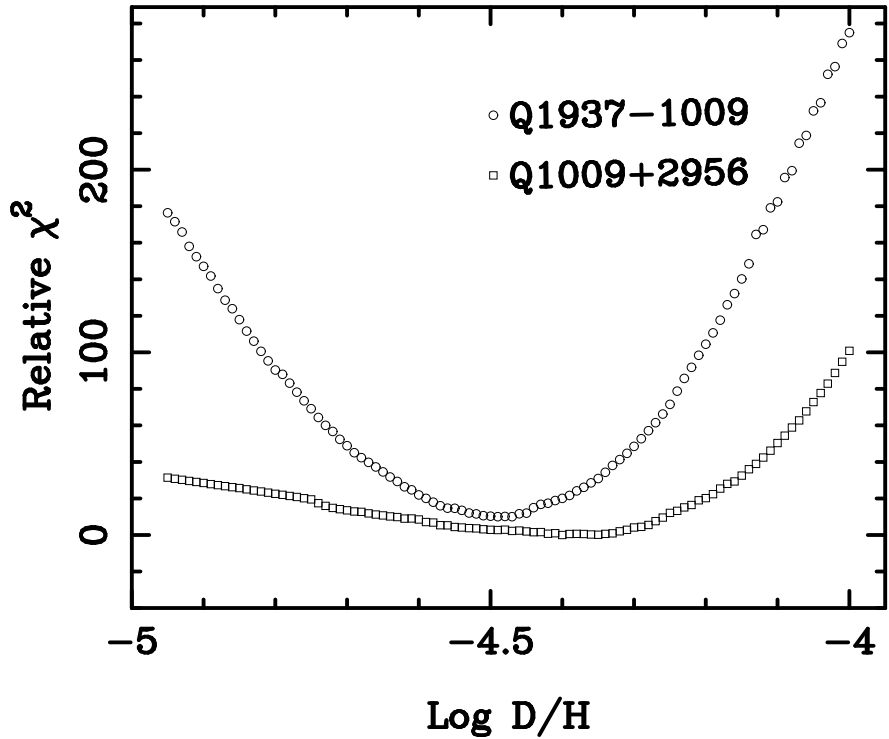,width=\textwidth}
\caption{$\chi^2$ functions of D/H in two separate absorption systems.  The
Relative $\chi^2$ is shown to facilitate a direct comparison.  The minima
represent the most likely values of D/H in each system, and $\Delta \,
\chi^2 = 4.0$ represents 95\% confidence limits.}
\end{figure} 

\begin{figure}
\psfig{file=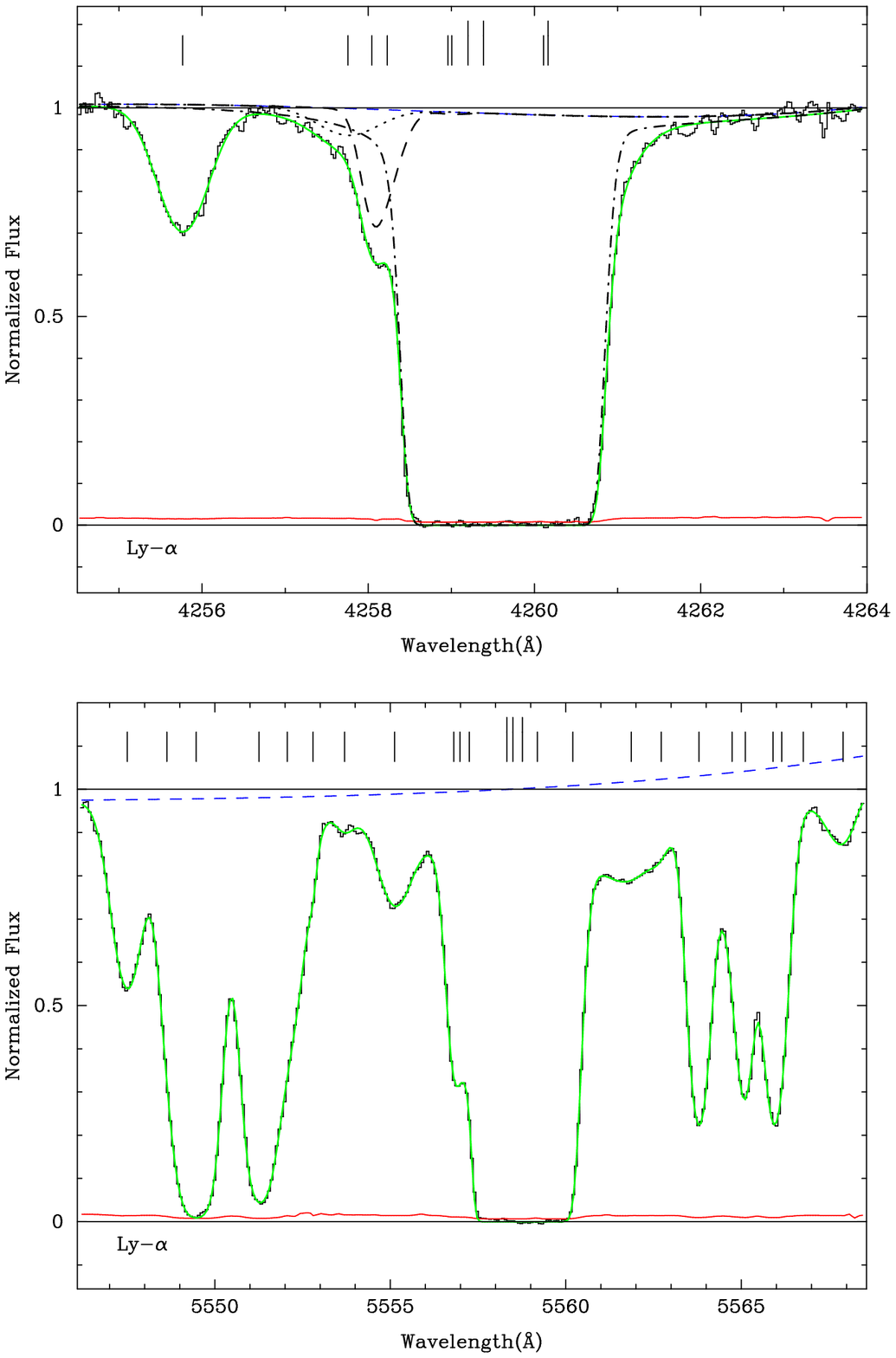,width=\textwidth}
\caption{ 
{\it Top}: \Lya region of $z=2.504$ system towards Q1009+2956. 
The histogram shows the pixels of the HIRES spectrum normalized to the
initial continuum estimate (solid line at unity).  The 1$\sigma$ errors
per pixel is shown as the solid line near zero.  The smooth dashed line
near unity is the best fit continuum with five degrees of freedom.  The
model profile is composed of the main H~I (dot-dashed) and D~I (dashed), 
and contaminating H~I (dotted). 
\protect\\
{\it Bottom}: \Lya region of $z=3.572$ system towards Q1937--1009.
The dashed line shows the best fit continuum level.}
\end{figure} 

The goal is to extract from the spectrum of a QSO absorption system 
the most likely value of D/H, and the confidence
intervals about that value.  We are still required to 
model the D/H absorption system (DHAS) with a finite set of parameters 
(based on physical assumptions) to compare to the observed spectrum.
With a given model, we can quickly minimize the difference between 
model and data, parameterized as $\chi^2$, and use the final parameters
to find the most likely value of D/H.  But the uncertainty in D/H depends
on correlations in the parameters, and is non-trivial to extract from the
best fit solution.  

The $\chi^2$ function depends on the N parameters included to
model the observed spectrum.  Instead of mapping the $\chi^2$ function in 
the full N dimensional parameter space, we choose to map $\chi^2$ versus
a single parameter, D/H (known as the ``ridge'' method).  
Towards Q1937-1009, for example, we compare a model with
234 free parameters to regions of the high-resolution spectrum which total
1298 pixels.  We note the inventory of free parameters:  3 main components 
in Lyman limit system with 4 parameters each (N(H~I), $z_{abs}$, T, $b_{tur}$)
+ 9 spectral regions with a total of 29 parameters for the continuum +
64 other H~I lines with 3 parameters each (N(H~I), $z_{abs}$, $b_{total}$)
which are near and affect the absorption lines of the D/H system + 1 parameter
for D/H: 3 x 4 + 29 + 64 x 3 + 1 = 234.  To construct the minimized $\chi^2$
function versus D/H alone, we choose a value for D/H, fix D/H for the
iteration, and allow the other 233 parameters to freely change to minimize
$\chi^2$.  We choose and fix a new value for D/H, and minimize $\chi^2$ again.
The results for both Q1937--1009 (this example) and Q1009+2956 are shown
in Fig. 3.  In Fig. 4 (bottom panel), note the large number (19) of H~I lines
which must be included in the \Lya region of Q1937-1009.

In this paper we present the difficulties  inherent to
measurements of D/H in QSO absorption systems, and discuss techniques we
have implemented to overcome the systematic effects.  
We outline an improved method to
measure D/H, which accounts for the total uncertainty, in contrast to the
statistically uncorrelated errors of D and H column densities summed
in quadrature.  We show that the best two current measurements of
D/H are complementary and give a robust measure of the primordial
value, $\rm{D/H} = 3.4 \pm \, 0.25 \times 10^{-5}$. 

\section{Systematic uncertainties in D/H measurements}

In this section, we describe the systematic uncertainties which are
known to exist in measurements of D/H in QSO absorption systems.
We explain how we account for the systematics, with the assumption that the
nature of the effect is understood.  

\subsection{Intrinsic QSO continuum}

The observed QSO spectra are combinations of the intrinsic 
spectrum of the source with intervening absorption features.
In the measurement of D/H, we are solely interested in the intervening
features.  We attempt and {\it require} the removal of the intrinsic
QSO spectrum (continuum) to analyze the absorption features in the observed
spectrum.  The QSO continuum changes on much longer wavelength scales than
the narrow absorption lines.  A polynomial of low-order traces the
smooth variations of the QSO continuum  and gives the
normalization of the observed spectrum to the QSO continuum.  
This procedure (called ``continuum fitting'') is not exact, 
and even worse, not well defined.  However, abundance measurements
in QSO absorption systems have not taken into account 
the uncertainties introduced by
subjectively fitting a fixed continuum to the observed spectrum.
Once a continuum fit is complete, the solution is assumed, for all practical
purposes, correct with no uncertainty.  We have overcome this problem
by assuming the continuum is not fixed or well determined.  As a part
of the abundance measurements, we include free parameters to account for the
unknown shape and level of the continuum in regions of interest in the 
observed spectrum.  We still fit an initial continuum (a
first guess), but this is not assumed absolute or correct, only a likely
estimate.  The procedure has been outlined in Burles \& Tytler (1997b),
and should be adopted for all abundance measurements 
which could be influenced   by continuum uncertainties.

As examples, we show the \Lya regions of Q1009+2956 (top)
and Q1937--1009 (bottom) in Fig 4.  
The initial continuum estimate is the solid line at unity and is
used to normalize the spectrum before profile fitting.  The
smooth dashed line near unity is the resulting best fit continuum
with five free parameters.  The level of the final continuum lies
near the initial estimate on the D/H \Lya feature, 
which confirms that the initial estimate was fairly
well chosen.  Allowing for a free continuum does not
necessarily  change the central value, but ensures that the
final uncertainties take properly into account the
uncertain nature of the intrinsic QSO continuum.

\subsection{Hydrogen Contamination of Deuterium}

The absorption features of the Lyman series of H~I and D~I are 
modeled to measure D/H in QSO absorption systems.  These features
reside exclusively in the \Lya forest and are subject to blending,
and therefore, contamination from other Lyman absorption.
The line density of the \Lya forest rises steeply with redshift,
so measurements at higher redshift, $z > 2$,  
are subject to more contamination
from random, unrelated \Lya lines.  The
{\it a posteriori} probability of unrelated H~I contamination 
can be estimated from the statistical distirbution of \Lya forest lines.
(Steigman 1994, Tytler \etal 1996, Tytler \& Burles 1997, Jedamzik
\& Fuller 1997, Hogan 1998).
On the other hand, structure
in the absorption systems will increase with decreasing redshift,
so contamination of measurements at lower redshift, $z \leq 1$, is
dominated by Lyman absorption correlated on scales of the D-H velocity
separation, 82 \kms.  We must test for the presence of contamination
in all measurements and ensure that the final results properly 
account for its existence.  Contamination always gives lower values
of D/H.  A measurement which does not include contamination in the
analysis gives an upper limit on D/H.  
In Fig. 4, we show the individual profiles of H~I (dot-dashed) and D~I
(dashed) in the Lyman limit system and unrelated H~I contamination redward of
the D~I profile (dotted line).  Contamination was included in analysis
towards both Q1937--1009 and Q1009+2956.

\subsection{Saturated Absorption features}

In measurements of D/H, one must select systems with high neutral hydrogen
column densities, log N(H~I) $> 17$, to be sensitive to D/H values
less than 10$^{-4}$.  In these systems, most of the lines of the Lyman
series have saturated (optical depth, $\tau > 3$) line centers.  
We obtain spectra with high signal-to-noise ratio (SNR) of the region
shortward of the Lyman limit to independently constrain N(H~I).
It is possible to obtain a direct measurement of the total H~I column
in systems with 17 $<$ log N(H~I) $<$ 18, 0.6 $< \tau_{LL}
<$ 6.0, where $\tau_{LL}$ is the optical depth of Lyman continuum absorption
at 912 \AA~ (rest frame).  
The residual flux shortward of the Lyman limit provides a
direct estimate of N(H~I), independent of other characteristics of the
system (e.g. velocity dispersion or deuterium column density).
To obtain an accurate measurement, we must account for
unrelated  Lyman absorption from other systems and the
uncertainty of the intrinsic QSO continuum near the Lyman limit (see sec. 4.1).
The majority of the Lyman absorption features can be
specified by their related absorption features in high-resolution spectra
(at higher wavelengths).  The remaining features are statistically
included from well determined \Lya forest distributions \cite{kir97}.
The QSO continuum near the Lyman limit can be determined to better
than 10\% by utilizing high-resolution spectra (c.f. Burles \& Tytler 1997a).
The uncertainty in the N(H~I) measurement is proportional to the
continuum uncertainty, $\Delta$ N(H~I) $= \Delta$ (cont) $/ \tau_{LL}$.
In Fig. 2, we show the Lyman limit region of the $z=2.504$ D/H system in
two separate spectra of Q1009+2956.  The high resolution spectrum
(top, FWHM = 8 \kms, 0.09 \AA) shows the resolved high order Lyman
lines up to Lyman-22.  The Lyman lines become unsaturated at the highest
orders, and give an optical depth in the continuum, $\tau_{LL} = 1.3$.
The low resolution spectrum (bottom, FWHM = 4 \AA) has better sensitivity
and shows significant flux shortward of the Lyman limit at 3200 \AA.

A one-parameter model, N(H~I), is compared with the low resolution
spectra to find the most probable value and confidence intervals of N(H~I)
(Burles \& Tytler 1997a).  The features seen in the model fit
in Fig. 2 are higher-order Lyman lines (\Lyb and above) associated 
with Lyman lines measured at longer wavelengths in the high resolution
spectrum.
\Lya absorption within $1.57 < z < 1.64$ is drawn from a distribution and 
included in the Monte Carlo analysis which gives the maximum likelihood of
N(H~I):   
log N(H~I) = 17.39 $\pm \, 0.06$, which includes 
the uncertainties from extrapolation of the unabsorbed QSO continuum.

\subsection{The Case of Mesoturbulence}

Levshakov \etal (1998, and references therein) have studied the effects
of correlations in turbulent velocity fields (mesoturbulence),
on line formation in the 
\Lya forest.  The standard line profiles (Voigt) are, in fact, the special
case of mesoturbulence when the correlation length goes to zero.
A substantial amount of work has been published to understand the impact
of mesoturbulence on the abundance measurements
in absorption systems, with emphasis on D/H  (Levshakov \& Kegel 1997, 
Levshakov \etal 1997,
Levshakov \etal 1998).  We would like to point out a few observational
constraints on mesoturbulence, and the limited role it plays (in our
view) in measurements of D/H.

(1) The turbulent component should be consistent with all lines in the
absorption system.  Mesoturbulent profiles will significantly differ from
Voigt profiles only when the turbulent component is comparable or larger
than the thermal component.  Both components contribute to the total 
velocity dispersion ($b_{total}$) which is measured in high resolution spectra:
$b_{total}^2 = b_{meas}^2 = b_{turb}^2 + b_{therm}^2.$ 
Comparisons of the metal line widths (e.g. C, Si) to the
hydrogen and deuterium line widths place strict limits on 
the turbulent component.  The heavy metal lines (Si and above) are
dominated by turbulence at temperatures $T < 10^5$ K and their
velocity dispersion usually gives a good measure of the 
turbulence in the system.  In the two systems we have measured D/H,
towards Q1937-1009 and Q1009+2956, the metal lines are very narrow,
which give very low velocity dispersions, $b(\rm{Si}) <6$ \kms.
The hydrogen lines are much wider, with typical widths 
$b(\rm{H}) \approx 20$ \kms, which shows 
that thermal, uncorrelated, line formation
is dominant in the optically thick Lyman lines.

(2) The line profiles of mesoturbulence can be distinguished
from Voigt profiles by fitting the entire Lyman series 
(Levshakov \& Kegel 1996).  The requirement for invoking mesoturbulence
should be determined by the observational data.  If the Lyman series
lines show evidence for mesoturbulent profiles, then one should take it 
into account. In D/H systems to date,
simple Voigt profiles can explain all of the profiles in the
Lyman series and do not require the mesoturbulent model.

(3) Mesoturbulence is likely to play an important role in a wide variety
of absorption line studies.  The hydrogen lines in QSO spectra which do
show deuterium belong to a class of absorbers with the narrowest intrinsic
widths, and are therefore subject
to the least amount of turbulent broadening.  On the other hand,
optically thick absorption lines of heavy metals (for instance, in
damped-\Lya systems) are prime candidates for studies of mesoturbulence. 

\subsection{Line blending}

The D/H absorption systems do not have simple, symmetric profiles
which can be modeled as a single component.  Analysis of the absorption
profiles of the Lyman lines and associated metal lines requires
multiple components.  The components are usually separated by less than
the intrinsic line widths of the individual components, which gives
rise to blended profiles of the Lyman series.  The severe blending
represents a loss of information.  The parameters describing
individual components are poorly constrained, but those 
describing the entire system can be characterized and measured.
Here lies the problem: the set of parameters describing the multiple
components of the D/H system are not orthogonal when the components
are blended.  The uncertainties in the parameters are strongly 
correlated, and the variance in the parameters does not represent the
total uncertainty.  Therefore, one cannot measure the
individual column densities of D and H accurately in each 
absorption component.

\acknowledgements{We are extremely grateful to the W. M. Keck foundation
which made this work possible.  We would like to thank our hosts at ISSI
  in Bern, J. Geiss and R. von Steiger, for their hospitality, 
and the editors of these proceedings, N. Prantzos, M. Tosi,
and R. von Steiger for their help.}


\begin{thebibliography}{99}
  
 
\bibitem[Burles \& Tytler 1997a]{bur97a}
Burles, S., \& Tytler, D. 1997, AJ, 114, 1330
 
\bibitem[Burles \& Tytler 1997b]{bur97b}
Burles, S., \& Tytler, D. 1997, ApJ, in press, astro-ph/9712108
 
\bibitem[Burles \& Tytler 1997c]{bur97c}
Burles, S., \& Tytler, D. 1997, submitted to ApJ, astro-ph/9712109

\bibitem[Cardall \& Fuller 1996]{card96}
Cardall, C. Y. \& Fuller, G. M. 1996, ApJ, 472, 435

\bibitem[Carswell et al. 1994]{car94}
Carswell, R. F., Rauch, M., Weymann, R. J., Cooke, A. J. \& Webb, J. K.
1994, MNRAS, 268, L1
 
\bibitem[Carswell et al. 1996]{car96}
Carswell, R. F., Webb, J. K., Lanzetta, K. M., Baldwin, J. A., Cooke, A. J.,
Williger, G. M., Rauch, M., Irwin, M. J., Robertson, J. G., \& Shaver, P. A.
1996, MNRAS, 278, 506
 
\bibitem[Copi et al. 1995]{cop95}
Copi, C.J., Schramm, D.N. \& Turner, M.S. 1995, Science, 267, 192

\bibitem[Fuller \& Cardall 1996]{ful96}
Fuller, G. M., \& Cardall, C. Y. 1996, Nucl. Phys. B, 51, 71

\bibitem[Hata et al. 1997]{hat97}
Hata, N., Steigman, G., Bludman, S., \& Langacker, P. 1997, Phys. Rev. D,
55, 540

\bibitem[Hogan 1998]{hog98}
Hogan, C. J. 1998, this volume

\bibitem[Jedamzik \& Fuller 1997]{jed97b}
Jedamzik, K. \& Fuller, G. M. 1997, ApJ, 483, 560
 
\bibitem[Kirkman \& Tytler 1997]{kir97}
Kirkman, D. \& Tytler, D. 1997, ApJ, 484, 848

\bibitem[Levshakov \& Kegel 1996]{lev96}
Levshakov, S. A., \& Kegel, W. H. 1996, MNRAS, 278, 497
  
\bibitem[Levshakov \& Takahara 1996]{lev96b}
Levshakov, S. A., \& Takahara, F. 1996, MNRAS, 279, 651
  
\bibitem[Levshakov \& Kegel 1997]{lev97}
Levshakov, S. A., \& Kegel, W. H. 1997, MNRAS, 288, 787

\bibitem[Levshakov \etal 1997]{lev97e}
Levshakov, S. A., Kegel, W. H., Takahara, F. 1997, submitted to MNRAS

\bibitem[Levshakov \etal 1998]{lev98}
Levshakov, S. A., Kegel, W. H., Takahara, F. 1998, this volume

\bibitem[Oke \etal 1995]{oke95}
Oke, J. B., Cohen, J. G., Carr, M., Cromer, J., 
Dingizian, A., Harris, F. H., Labrecque, S., Lucinio, R., 
Schaal, W., Epps, H., Miller, J. 1995, PASP, 107, 375

\bibitem[Rugers \& Hogan 1996a]{rug96a}
Rugers, M., \& Hogan, C. 1996, ApJ, 459, L1
 
\bibitem[Rugers \& Hogan 1996b]{rug96b}
Rugers, M., \& Hogan, C. 1996, AJ, 111, 2135

\bibitem[Sarkar 1996]{sar96}
Sarkar, S. 1996, Rep. Prog. Phys., 59, 1493

\bibitem[Schramm \& Turner 1997]{sch97}
Schramm, D. N. \& Turner, M. S. 1997, submitted to Rev. Mod. Phys.
 
\bibitem[Smith \etal 1993]{smi93}
Smith , M. S., Kawano, L. H. \& Malaney, R. A. 1993, ApJS, 85, 219
 
\bibitem[Songaila et al. 1994]{son94}
Songaila. A., Cowie, L. L., Hogan C. J. \& Rugers, M. 1994, Nature, 368, 599

\bibitem[Songaila et al. 1997]{son97}
Songaila, A., Wampler, E. J., \& Cowie, L. L. 1997, Nature, 385, 137

\bibitem[Steigman 1994]{ste94}
Steigman, G. 1994, MNRAS, 269, L53

\bibitem[Tytler et al. 1996]{tyt96}
Tytler, D., Fan, X-M., \& Burles, S. 1996, 381, 207 

\bibitem[Tytler \& Burles 1997]{tyt97}
Tytler, D., \& Burles, S. 1997, in "Origin of Matter and
Evolution of Galaxies", eds. T. Kajino, Y. Yoshii \& S. Kubono
(World Scientific Publ. Co.: Singapore), 37

\bibitem[Vogt \etal 1994]{vog94}
Vogt, S. \etal 1994, Proc. SPIE, 2198, 362
\bibitem[Walker \etal 1991]{wal91}
Walker, T. P., Steigman, G., Schramm, D. N., Olive, K. A. \& Kang, H. S. 1991
ApJ, 376, 51

\bibitem[Wampler 1996]{wam96}
Wampler, E. J. 1996, Nature, 383, 308

\bibitem[Wampler et al. 1996]{wam96}
Wampler, E.J., Williger, G.M., Baldwin, J.A., Carswell, R.F.,
Hazard, C. \& Mcmahon, R.G. 1996, A\&A, 316, 33
 
\bibitem[Webb et al. 1997]{web97}
Webb, J. K., Carswell, R. F., Lanzetta, K. M., Ferlet, R., Lemoine, M.,
Vidal-Madjar, A., \& Bowen, D. V. 1997, Nature, 388, 250
 
\end{thebibliography}
\end{document}